# Statistical Dimension Identification and Implementation for Student Progression System

**Harkiran Kaur**
Department of Computer Science and Engineering, Thapar Institute of Engineering and Technology,
(Deemed to be University, Patiala),India
**Aanchal Phutela**
Department of Computer Science and Engineering, Thapar Institute of Engineering and Technology,
(Deemed to be University, Patiala),India

**Abstract-** Descriptive Analytics is the summarization of the past data and generates some useful patterns from that data. This work focuses on analyzing and querying large academic dataset for generating Student Progression using visualization and dashboards. Presently projects on Progression Systems exist but no descriptive or predictive analytics has been performed on these datasets. The proposed system collects data from different departments of University, store data into the large data warehouse of the University and generate validated set of KPIs, based on the past dataset of University's department. These KPIs are obtained after applying Statistical techniques on various sets of dimension in the academic datasets. After completion of this step, analysis of the data has been achieved with Online Analytical Processing (OLAP) operations, which have been show cased with the help of visualization and dashboards.

## 1. INTRODUCTION

Descriptive Analytics as the name suggests, 'describe' or summarize the data into some useful information and possibly formulate the data for further analysis that is comprehensible by humans. Some common techniques engaged in Descriptive Analytics are observations, case studies, and surveys. In the proposed work the given dataset is analyzed by applying statistical methods on the academic dataset using IBM SPSS Statistics tool. The main agenda of this study is to create student progression system. For this purpose, the descriptive analytics must be performed on only those features of the academic dataset which have a huge impact on the projected goal and these features are commonly called Key Performance Indicators (KPI's).

In its simplest form, a KPI is a type of measurement that helps you to recognize how your organization or department is carrying out. A good KPI will help you and your team to recognize whether you're choosing the right path in the direction of your planned goals or not. A KPI must be effective if it should follow the SMART criteria. SMART refers to Specific, Measurable, Achievable, Relevant, and Time-bound [1].

The present study perform analytics on academic dataset of students in a University, which can be done by descriptive analytics (the introductory stage of the data analysis) and creates the summary of historical data and generate some useful information from the past trends in the dataset.

The vital success of an organization is its ability to analyze the data, find some major facts in it and take some actions towards the changes. The task of finding major facts in these dataset is performed by validated selection of KPI. Further the corresponding actions are taken as per the data analytic technique applied on these KPIs contained in the dataset.

## 2. KPI CLASSIFICATION AND IDENTIFICATION METHOD

The major issue in KPIs identification is that, there are N number of KPIs to choose from. If we are choosing the wide of the mark, then we are calculating something that doesn't line up with our objectives. Every organization needs to analyze the results for their respective business problems and consider just those set of KPIs which are most relevant for an organization's further monitoring, evaluating, planning and decision making. Therefore, it is a prerequisite to identify a set of KPIs for a specific organization for its mission, vision and values on the subject of an organization's approach. A KPI should be: relevant, realistic, specific, attainable, measurable, and used to identify trends, timely, understood, agreed, reported, governed or resourced [1].

There are several methods and techniques you can use for selecting subset of features which helps your model to perform better and efficiently. These include: Pearson's Correlation, Linear Discriminant Analysis (LDA), ANOVA and Chi-Square Test. These statistical filter methods are used as a preprocessing step for identifying KPIs. In this the selection feature is independent from any machine learning algorithm. Fig 1 describes the implementation of filter methods.

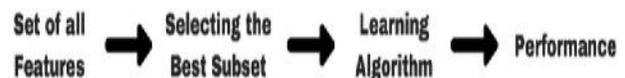

**Fig.1. Filter Methods [4]**

Some other methods include: Forward Selection, Backward Elimination and Recursive Feature elimination, these are Wrapper Methods. Figure 2 demonstrates the processing of wrapper methods, in which learning algorithms are used for the selection of best subset of features [4].

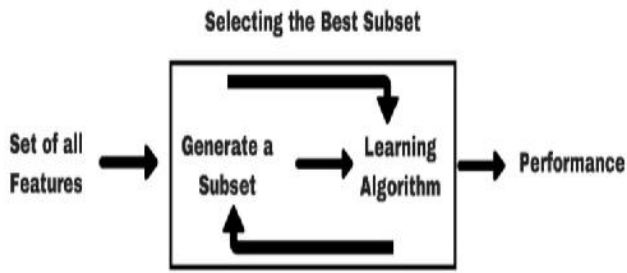

**Fig. 2. Wrapper Methods [4]**

There are also some mehods which combine the qualities' of Filter and Wrapper Methods that are called Embedded Methods which includes LASSO and RIDGE Regression Methods. Figure 3 illustsrates the processing of embedded methods. In which the best subset of features are selected from set of all features.

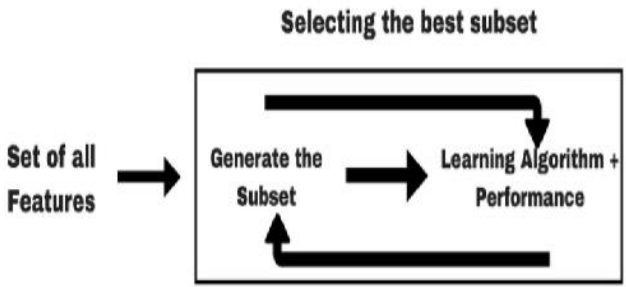

**Fig. 3. Embedded Methods [4]**

### 3. RELATED WORK

Ratapol Wudhikarn et al. [10] proposed a work for intellectual capital management for which the author used the Delphi Method. Mainly, the research has been focused on the identification of important indicators, which were significant for the business performance. Since the research has been done on Delphi Method, it will help the organization to manage the intellectual capital in the business logistics more effectively. This method is better than the completely depending upon expert's advice and also helped in determining important KPIs in business logistics.

Key performance indicator (KPI) framework proposed by Sangeeta M. Joshi [3] can be used by an educational institute to meet the developmental necessities across the world. In this work, the authors have developed a performance management system (PMS). Firstly the authors categorizes the KPIs in different areas and then evaluate performance of faculty by identifying the KPIs. The PMS will help to produce faculty ranking and also helps the institutes to rise their quality standards.

Dwi Al Aji Suseno et al. [9] proposed a work of determining bonus in enterprise resource planning using KPI. In this work, the authors define how to calculate bonus by calculating the weight and analyzing the KPIs. In general, there are many ways to determine the bonus in the company but it leads to some drawbacks for the company. This is because there is no clear idea that determines the significant factors for the organization. SMART criteria had been used by the authors for analysis and determining the bonus in the company. The authors concluded that, determining bonus depends upon KPI is more efficient than seniority-based distribution of bonus among the employees. Another observation was that, if the employee can achieve the KPI goal he will get more bonus.

According to Paulo Roberto Martins de Andrade et al. [8] having a system like Business Indicator Management helps to meet their needs in terms of information availability and agility as well. The proposed work gives the integrated approach to manage the KPIs, through which real-time information about organizations can be obtained and further the actual situation of the companies can be analyzed. This helps the organization to increase their productivity and also decision making efficiency.

Jinsoo Park et al. [6] proposed a work for manufacturing scheduling, it is almost unmanageable for schedulers to study all the constraints. So, the authors have proposed an approach of simulation-based advanced planning and scheduling (APS). In this work, the authors propose a new method that meet the features of any process by selecting appropriate KPIs. The authors verify this method with empirical analysis whether they meet the requirements of KPIs or not. In which the authors associate the outputs derived from simulation-based APS with modifications of domain experts and propose a framework to identify the appropriate KPIs.

David Plandor et al. [7] proposed an article, in which the authors invent an application for producing KPIs. First of all, the authors initiated with the results of the company's business diagnostic process. Formerly, the results was ready for the KPIs analysis, which was stored in the database. The author's objective is to create a set of KPIs, for which they used genetic algorithm. The author developed a brand new software application as a tool for KPI collection.

Worarat Krathu et al. [5] proposed an analysis for inter-organizational relations (IOR's) that are important for collaborations between businesses. In this work, the authors proposed KPIS for measuring success factors. For this work, the authors presented a method for identifying IOR's. The proposed methods takes the semantics and data types of both data elements come to precise results. The authors applied this technique on real-world industry MIGs and offered a set of inter-organizational KPIs. The KPIs offered can be used for the evaluation of IOR's.

### 4. PROPOSED WORK

The proposed work applies statistical filter methods: ANOVA test and correlation testing on academic dataset, to obtain validated set of KPIs.

In this research, authors of this paper used two steps for creating Student Progression System. These steps are:

**STEP 1: Evaluation of applied statistical technique and select the KPIs**

This step further involve following sub steps.

The steps are: a) define candidate KPIs, b) define null hypothesis and alternate hypothesis for the selected candidate KPIs in first step, c) perform descriptive analytics to identify their correlation, d) formulation of condense list of KPIs.

**Step a. Define candidate KPIs**

This study utilizes Academic Dataset of the students from the University Database which includes different factors of students like number of backlogs, number of semesters, regularity, Extra curriculum activities, projects done, research work done, their grades or CGPA or many other dimensions. These are the candidate KPIs for the selected domain.

**TABLE I CATEGORY-WISE CANDIDATE KPIs.**

| Sr. No. | Candidate KPIs | Category |
|---|---|---|
| 1. | No. of Backlogs | Quantitative KPI |
| 2. | Extra curriculum activities | Leading KPI |
| 3. | Regularity | Actionable KPI |
| 4. | CGPA | Outcome KPI |
| 5. | State | Quantitative KPI |
| 6. | Projects | Quantitative KPI |
| 7. | Research Work | Qualitative and Quantitative KPI |
| 8. | All Rounder Score | Leading KPI |
| 9. | Number of Semester in the course | Quantitative KPI |
| 10. | Number of Subjects | Quantitative KPI |

**Step b. Define Null Hypothesis (H0) and Alternate Hypothesis (H1) for the selected candidate KPIs in first step**

In the proposed work we are using filter statistical techniques for feature selection (significant factors), including ANOVA, Correlation Test and from the list of given factors. These tests are applied on various combination of candidate KPIs. Some of them have been showcased in the coming paragraphs. Let us take the Null and Alternate Hypothesis for two factors that is:
H0: First factor have no significant effect on Second factor.
H1: First factor have significant effect on Second factor.
Where H0 represents the Null Hypothesis and H1 represents the Alternate Hypothesis.
These tests has been conducted under 0.05 significant level.
a. ANOVA Test
One-way Analysis Of Variance(ANOVA) will be performed on these factors, which will determine whether there is any statistically significant differences between the means of two or more independent (unrelated) groups or not. If the results would be less than the p-value then the null hypothesis will be rejected and alternate hypothesis will be accepted and vice- versa. ANOVA test has been performed on several pair of factors as shown below:

**Set 1: Regularity and Extra Curriculum Activities**
The corresponding hypothesis would be:
H0: Regularity have no significant effect on extra curriculum activities.
H1: Regularity have significant effect on extra curriculum activities.

**ANOVA**

Regularity

| | Sum of Squares | df | Mean Square | F | Sig. |
|---|---|---|---|---|---|
| Between Groups | 10.085 | 2 | 5.042 | 19.433 | .000 |
| Within Groups | 12.195 | 47 | .259 | | |
| Total | 22.280 | 49 | | | |

**Fig.4. Results of ANOVA Test**

Now, according to the Fig 4, the results shows that p-value is less than 0.05 then, the null hypothesis will be rejected and the alternate hypothesis will be accepted. It means that Regularity has significant effect on Extra Curriculum Activities.

**Set 2: Regularity and CGPA of students.**
The corresponding hypothesis for this set them would be
H0: Regularity have no significant effect on CGPA of students.
H1: Regularity have significant effect on CGPA of students.

**ANOVA**

CGPA

| | Sum of Squares | df | Mean Square | F | Sig. |
|---|---|---|---|---|---|
| Between Groups | 17.189 | 3 | 5.730 | 12.860 | .000 |
| Within Groups | 20.494 | 46 | .446 | | |
| Total | 37.683 | 49 | | | |

**Fig.5. Results of ANOVA Test**

Now, according to the Fig 5, the results shows that p-value is less than 0.05 then, the Null Hypothesis will be rejected and the Alternate Hypothesis will be accepted. It means that Regularity have significant effect on CGPA.

**Set 3: Number of semester in the course and CGPA if students.**
The corresponding hypothesis for this set would be:
H0: Number of semester have no significant effect on CGPA of students.
H1: Number of semester have significant effect on CGPA of student.

**ANOVA**

CGPA

| | Sum of Squares | df | Mean Square | F | Sig. |
|---|---|---|---|---|---|
| Between Groups | .220 | 2 | .110 | .138 | .871 |
| Within Groups | 37.463 | 47 | .797 | | |
| Total | 37.683 | 49 | | | |

**Fig.6. Results of ANOVA Test**

Now, according to the Fig 6, the results shows that p-value is greater than 0.05 then, the Null Hypothesis will be accepted and the Alternate Hypothesis will be rejected. It means that Number of Semester in a Course has no significant effect on CGPA of students.

b. Correlation Test

The authors used two factors at a time and explored the correlation between them, as the value of correlation test tells about the dependency of factors upon each other. If the correlation test will give the positive value then it means factors are directly proportional with each other and if its value is negative then it means factors are inversely proportional with each other. In IBM SPSS tool, we the authors have implemented this technique on following set of candidate KPIs.

**Set 1: Regularity and Extra Curriculum Activities.**

In Fig 7, negative correlation between these factors will be observed which identifies the inversely proportional relationship between them. It means that when one parameter increases other will decreases and vice-versa.

**Fig.7. Results of Correlation Test**

**Set 2: Regularity and CGPA.**

In Fig 8, positive correlation between these factors has been observed which identifies the directly proportional relationship between them. It means that when regularity increases CGPA also increases and vice-versa is also true.

**Fig.8. Results of Correlation Test**

**Set 3: Number of Semester in a course and CGPA.**

In figure 9, positive correlation between these factors has been observed which identifies the directly proportional relationship between them. It means that when number of semester increases CGPA will decreases and vice-versa. But the figure also shows that the significant value if greater than 0.05 which identifies there is no significance of number of semesters on the CGPA of the students.

**Fig.9. Results of Correlation Test**

## 5. RESULTS AND FINDINGS

**TABLE II CONDENSED LIST OF KPIs**

| S. No. | Condense KPIs | Category |
|---|---|---|
| 1. | No. of Backlogs | Quantitate KPI |
| 2. | Extra curriculum activities | Leading KPI |
| 3. | Regularity | Actionable KPI |
| 4. | CGPA | Outcome KPI |
| 5. | Projects | Quantitative KPI |
| 6. | Research Work | Qualitative and Quantitative KPI |
| 7. | All Rounder Score | Leading KPI |

In Table 2 describes the condense list of KPIs retrieved after applying step1 on set of candidate KPIs. By applying statistical methods on them. As the results shows some KPIs are significant for our goal so we will apply the descriptive techniques only on the condense list of the KPIs which will increase the performance of the system and make the decision making process easier.

## 6. VISUALIZATION

In Fig10, the authors have presented some visualization on dashboards. In this figure the authors have shown that, the changes occurred in the dataset will reflect directly on the charts as shown in the figure. The authors have used a slicer in which slicing of data has been done and the changes were immediately reflects on the dashboards. As shown when the field M.tech from the slicer has chosen the change on resultant dataset occurred and it reflects on the visualization. In this way, we can make our data more visible and easier to understand for analysis.

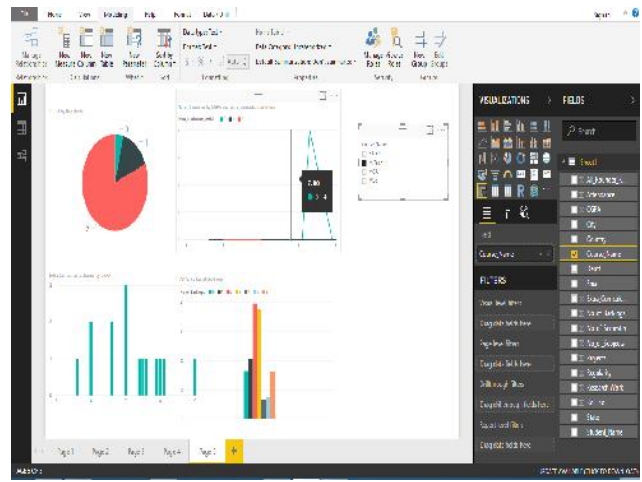

**Fig.10. Visualization of Academic KPIs**

## 7. CONCLUSION

In the proposed work, the authors have implemented the Statistical techniques such as Hypothesis Testing and Correlation testing to generate the optimized and most significant set of KPIs for the Student Progression System. Further, descriptive analytics has been applied on the Academic dataset based on the obtained list of KPIs. This descriptive analysis is demonstrated with the help of visualization of KPIs on the Academic Dataset on the dashboards. These visualizations supports the course wise analysis of student progressions, based on the slicer feature of OLAP. This work can be further extended by applying more of the OLAP operations on these visualizations and generate multidimensional data views from the Academic dataset.